\documentclass[onecolumn, 12pt]{IEEEtran}
\newtheorem{Lemma}{Lemma}
\newtheorem{Theorem}{Theorem}
\newtheorem{Remark}{Remark}
\usepackage[caption=false,font=footnotesize]{subfig}
\usepackage{amssymb}
\usepackage{amsfonts}
\usepackage{color}
\usepackage{amsmath}%
\usepackage{graphicx}
\usepackage{bm}

\usepackage{url}
\usepackage{cite}
\usepackage{tabularx,colortbl}
\usepackage[T1]{fontenc}
\usepackage[colorlinks,linkcolor=black,anchorcolor=black,
citecolor=black]{hyperref}

\usepackage{setspace}

\begin{document}
	\title{ Distributed IRS with Statistical Passive Beamforming for MISO Communications}
		
\author{Yuwei~Gao,	Jindan~Xu,~\IEEEmembership{Student~Member,~IEEE,} Wei~Xu,~\IEEEmembership{Senior~Member,~IEEE,}			Derrick~Wing~Kwan~Ng,~\IEEEmembership{ Senior~Member,~IEEE}				and~Mohamed-Slim~Alouini,~\IEEEmembership{ Fellow,~IEEE}
}

	\maketitle
	
	\begin{abstract}
		 Intelligent reflecting surface (IRS) has recently been   identified as a prominent technology with the ability of enhancing wireless communication by dynamically manipulating the  propagation environment. This paper investigates a multiple-input single-output (MISO) system deploying distributed  IRSs. For practical considerations, we propose an efficient design of passive reflecting beamforming for the IRSs to exploit statistical channel state information (CSI) and analyze the achievable rate of the network taking into account the impact of CSI estimation error. The ergodic achievable rate is derived in a closed form, which provides insightful system design guidelines. Numerical results confirm the accuracy of the derived results and unveil the performance superiority of the proposed distributed IRS deployment over the conventional centralized deployment.
	\end{abstract}
	

	\IEEEpeerreviewmaketitle

\section{Introduction}
%
%
%
%

As a key technology of the fifth-generation (5G) wireless
networks,
the merits of massive MIMO are reaped at the cost of increased power consumption and hardware cost \cite{R}.
To circumvent these challenges, intelligent reflecting surface (IRS), which is enabled by recent developments in the  metamaterial technology,
has become a promising  alternative to enhance the performance of wireless communication systems by exploiting a passive antenna array\cite{xujindan}. 

IRS is a reprogrammable metasurface comprising of a large number of cost-effective passive reflecting  elements that can  dynamically  manipulate impinging electromagnetic waves, thereby constructing favorable wireless channels.  \textcolor{black}{The IRS can be easily attached to or removed from existing objects in the environment (e.g., walls and ceilings) to serve distant groups of users. As a result, recent studies have focused on the fundamental challenges in applying IRSs in practice, e.g., channel estimation, passive beamforming design, and performance evaluation.}
For instance, in \cite{passzhang}, the authors established a ``signal hotspot" as well as an ``interference-free zone"  by jointly optimizing the  reflecting beamformers and the deployment of multiple IRSs. Also, in \cite{yuhan}, it was proved that the introduction of an IRS can considerably enhance the  spectral efficiency of a large-scale antenna system even with a coarsely discretized reflecting beamformer equipped at the IRS. To fully unlock the potential of IRS, the design of pragmatic channel estimation methods were also investigated for IRS systems \cite{estzhang3}\cite{estimationzhanf}. \textcolor{black}{ In \cite{ref2}\cite{spatial}, the ergodic achievable rate was characterized averaged over both channel fading and random IRS/UE locations to investigate the system level performance aided by distributed IRSs.}  However, most of these studies focused on a centralized deployment of a single IRS, while	 practical limitations,
 such as CSI imperfection, have rarely been considered for the design of distributed IRS beamforming. More importantly, a thorough study on the  performance of distributed IRS systems has not been reported in the literature yet. 

 \textcolor{black}{In this paper, we consider a multiple-input single-output (MISO) system assisted  by multiple distributed IRSs which exploit statistical CSI, i.e., the statistical mean and correlation of the channel matrices, for the design of passive beamforming and the corresponding performance at the receiver is analyzed. } In particular, we first propose a low-complexity passive beamforming design for the IRSs which exploit only statistical correlation information of the channels. Then, we analyze the ergodic rate of the network with the considerations of CSI estimation error at the receiver. An exact closed-form expression is obtained for characterizing the ergodic rate.
 Furthermore, we  find that the ergodic rate is significantly improved with the increasing number of line-of-sight (LoS) paths between base station (BS)-to-IRS and IRS-to-user links. The superiority of the distributed IRSs over a centralized IRS is verified since the former increases the possibility of the presence of LoS
 channels. 

\textit{Notations:}	 $\mathbb{E}\{\cdot\}$ is the statistical expectation operation.  $[\mathbf{{H}}]_{i,j}$ is  the $(i,j)$th-element of matrix $\mathbf{{H}}$. Operators  tr($\cdot$), $\text{diag}\{\cdot\}$, $\text{blkdiag}\{\cdot\}$, and $\angle$ represent the trace, diagonalization, the block diagonalization of an input matrix, and angle respectively. Operator $\lVert\cdot\rVert$ is the Euclidean norm of an input vector.
Operator $\otimes$ is the Kronecker product.

\section{System Model}
\label{section2}

\textcolor{black}{We  focus on the case where a typical UE at any arbitrary location in the cell is served by multiple distributed IRSs on the user, compared with a larger-sized centralized IRS with the same total number of antennas on IRSs. 
} We consider a single-user  MISO downlink system which consists of one BS and $N$ IRSs in a single cell. The BS is equipped with  $M$ antennas and each IRS has $L$ passive reflectors. 
Assume that the channel between the BS and IRS $n$, $\forall n\in\{1,...,N\}$, follows a correlated Rician distribution, given as
\begin{equation}
\label{1}
\mathbf{H}_{n}=\sqrt{\frac{\beta_{n}K_{1,n}}{K_{1,n}+1}}\bar{\mathbf{H}}_{n}+\sqrt{\frac{\beta_{n}}{K_{1,n}+1}}{\widetilde{\mathbf{H}}}_{n},
\end{equation}
where $\beta_{n}$ is the large-scale path-loss, 
$\bar{\mathbf{{H}}}_{n}\in\mathbb{C}^{L\times M}$ is the LoS component, $\widetilde{\mathbf{H}}_{n}\sim \mathcal{CN}(\mathbf{0},\mathbf{R}_{n}\otimes\mathbf{I}_{M})$ denotes the  non-LoS (NLoS) component with spatial correlation, $\mathbf{R}_{n}\succeq\mathbf{0}$, and $K_{1,n}$ is the Rician $K$-factor. The entry $[\bar{\mathbf{{H}}}_{n}]_{i,j}=e^{-j2\pi\frac{ d_{n,ij}}{\lambda}}$, where $\lambda$ is the wave length and $d_{n,ij}$ is the distance between reflecting element $i$ of IRS $n$ and antenna $j$ of the BS.
Similarly, we  denote the channel between IRS $n$ and the user by
\begin{equation}
\label{2}
\mathbf{g}_{n}=\sqrt{\frac{\alpha_{n}K_{2,n}}{K_{2,n}+1}}\bar{\mathbf{g}}_{n}+\sqrt{\frac{\alpha_{n}}{K_{2,n}+1}}{\widetilde{\mathbf{g}}}_{n},
\end{equation}
where $\alpha_{n}$, $K_{2,n}$,   ${\bar{\mathbf{g}}}_{n}\in\mathbb{C}^{L\times 1}$, and $\widetilde{\mathbf{g}}_{n}\sim\mathcal{CN}(\mathbf{0},\mathbf{I}_{L})$ are similarly defined as in (\ref{1}).

The direct link between the BS and the user, denoted by $\mathbf{h}_{d}\in\mathbb{C}^{M\times 1}$, is assumed Rayleigh distributed with zero mean and unit variance. \textcolor{black}{For the case where a direct link between the BS and the user is available, the user may better communicate directly with the BS.
} Then, the received signal at the user is given by\footnote{We assume that the signals experienced more than one reflections are ignored as they are severely attenuated compared with the directly reflected one \cite{ref2}.}
\begin{equation}
\label{3}
{y}=\sqrt{P}\left(\mathbf{h}_{d}^{H}+\sum_{n=1}^{N}\mathbf{g}_{n}^{H}\boldsymbol{\Phi}_{n}\mathbf{H}_{n}\right)\mathbf{f}{s}+{w},
\end{equation}
where $\boldsymbol{\Phi}_{n}=\text{diag}\{\textit{e}^{j\phi_{n,1}},...,\textit{e}^{j\phi_{n,L}}\}$, $\phi_{n,l}\in[0, 2\pi)$, $\forall l\in\{1,...,L\}$, is the $l$th phase shift introduced by the reflecting beamforms of IRS $n$, ${P}$ is the transmit power of the BS, ${w}$ denotes the additive white Gaussian nosie (AWGN) with zero mean and variance $\sigma_{w}^{2}$, $\mathbf{f}\in\mathbb{C}^{M\times 1}$ is the beamforming vector adopted at the BS, and $s\in\mathbb{C}$ is the source symbol satisfying $\mathbb{E}\{|s|^{2}\}=1$.
 For notational simplicity, we define the stacked channel matrices as $\mathbf{H}=[\mathbf{H}_{1}^{H},...,\mathbf{H}_{N}^{H}]^{H}\triangleq\bar{\mathbf{{H}}}+\mathbf{K}_{1}\widetilde{\mathbf{H}}$, where $\bar{\mathbf{H}}=\left[\sqrt{\frac{\beta_{1}K_{1,1}}{K_{1,1}+1}}\bar{\mathbf{H}}_{1}^{H},...,\sqrt{\frac{\beta_{N}K_{1,N}}{K_{1,N}+1}}\bar{\mathbf{H}}_{N}^{H}\right]^{H}$ and $\widetilde{\mathbf{H}}=[\widetilde{\mathbf{H}}_{1}^{H},...,\widetilde{\mathbf{H}}_{N}^{H}]^{H}$. It is readily known that $\widetilde{\mathbf{H}}\sim\mathcal{CN}(\mathbf{0},\mathbf{R}\otimes\mathbf{I}_{M})$ with $\mathbf{R}=\text{blkdiag}\{\mathbf{R}_{1},...,\mathbf{R}_{N}\}$ and  $\mathbf{K}_{1}=\text{blkdiag}\left\{\sqrt{\frac{\beta_{1}}{K_{1,1}+1}}\mathbf{I}_{L},...,\sqrt{\frac{\beta_{N}}{K_{1,N}+1}}\mathbf{I}_{L}\right\}$. We also define  $\mathbf{G}\triangleq\bar{\mathbf{G}}+\mathbf{K}_{2}\widetilde{\mathbf{G}},$ where $\mathbf{K}_{2}=\text{blkdiag}\left\{\sqrt{\frac{\alpha_{1}}{K_{2,1}+1}}\mathbf{I}_{L},...,\sqrt{\frac{\alpha_{N}}{K_{2,N}+1}}\mathbf{I}_{L}\right\}$, $\bar{\mathbf{G}}=\text{blkdiag}\left\{\sqrt{\frac{\alpha_{1}K_{2,1}}{K_{2,1}+1}}\bar{\mathbf{g}}_{1}^{H},...,\sqrt{\frac{\alpha_{N}K_{2,N}}{K_{2,N}+1}}\bar{\mathbf{g}}_{N}^{H}\right\}$, and $\widetilde{\mathbf{G}}=\text{blkdiag}\{\widetilde{\mathbf{g}}_{1}^{H},...,\widetilde{\mathbf{g}}_{N}^{H}\}$. 

Then, let $\mathbf{v}_{n}\triangleq[\textit{e}^{j\phi_{n1}},...,\textit{e}^{j\phi_{nL}}]^{H}$, $\forall n\in\{1,...,N\}$, and $\mathbf{G}\triangleq\text{blkdiag}\{\mathbf{g}_{1}^{H},...,\mathbf{g}_{N}^{H}\}$. The received signal at the user in (\ref{3}) can be  equivalently rewritten as
\begin{equation}
\label{4}
{y}=\sqrt{P}\left(\mathbf{h}_{d}^{H}+\mathbf{v}^{H}\mathbf{Z}\right)\mathbf{f}{s}+{w},
\end{equation}
where $\mathbf{v}^{H}=[\mathbf{v}_{1}^{H},...,\mathbf{v}_{N}^{H}]$ and $\mathbf{Z}\triangleq\mathbf{GH}$. 

Now, we are ready to present the design of the reflecting beamformer, $\mathbf{v}$, and  charaterize its ergodic achievable rate.
\section{Distributed IRSs with Statistical CSI}
\label{section3}
\subsection{Distributed Reflection Design with Statistical CSI}
To facilitate the control of the distributed IRSs in practice, we consider that only statistical CSI is available for the design of the reflecting beamformers,  $\mathbf{v}$, of  the IRSs \cite{CSIMIMO} for  maximizing the ergodic achievable rate of the network. From (\ref{4}), the equivalent SNR of the received signal depends on the availability of CSI, which makes the problem challenging. In general, with limited system resources, the transmitter can acquire only imperfect CSI through a dedicated channel estimation.
 Consider the application of a typical linear minimum mean-squared error
   (LMMSE) channel estimate. It follows from the property of MMSE estimation, e.g., \cite{jinshi}\cite{CSIMIMO}, that  the CSI estimates of the channels,  ${\mathbf{Z}}$ and $\mathbf{h}_{d}$,  can be written as $\mathbf{Z}=\hat{\mathbf{Z}}+\mathbf{E}_{\mathrm{Z}}$ and   $\mathbf{h}_{d}=\hat{\mathbf{h}}_{d}+{\mathbf{e}}_{d}$, respectively, where $\mathbf{E}_{\mathrm{Z}}\sim \mathcal{CN}(\mathbf{0}, \xi\mathbf{I}_{NL}\otimes\mathbf{I}_{M})$ and ${\mathbf{e}}_{d}\sim \mathcal{CN}(\mathbf{0}, \xi\mathbf{I}_{M})$   are, respectively, the corresponding  estimation errors.  Letting $T$ be the  length of training sequences and $\rho$ as the
  signal-to-noise ratio (SNR) of the training sequence, we have $\xi=\frac{1}{1+T\rho}$ \cite{error}. 
  By exploiting the typical beamforming technique of maximum ratio  transmission (MRT) at the BS due to its optimality  maximizing the receive SNR in a single-user system, i.e., $\mathbf{f}=\frac{({\hat{\mathbf{h}}_{d}^{H}+\mathbf{v}^{H}\hat{\mathbf{Z}}})^{H}}{\left\|{\hat{\mathbf{h}}_{d}^{H}+\mathbf{v}^{H}\hat{\mathbf{Z}}}\right\|}$, the received signal in (\ref{4}) becomes
\begin{equation}
y=\left(\lVert\hat{\mathbf{h}}_{d}^{H}+\mathbf{v}^{H}{\hat{\mathbf{Z}}\rVert}+
\frac{({\mathbf{e}}_{d}^{H}+\mathbf{v}^{H}{\mathbf{E}_{\mathrm{Z}}})(\hat{\mathbf{h}}_{d}+{\hat{\mathbf{Z}}}^{H}\mathbf{v})}{\lVert\hat{\mathbf{h}}_{d}^{H}+\mathbf{v}^{H}{\hat{\mathbf{Z}}}\rVert}\right)\sqrt{P}s+w.
\end{equation}
Then, the system ergodic achievable rate can be expressed as 
\begin{align}
\label{6}
C	=\mathbb{E}\left\lbrace\text{log}_{2}\left(1+P\frac{\lVert\hat{\mathbf{h}}_{d}^{H}+\mathbf{v}^{H}{\hat{\mathbf{Z}}}\rVert^2}{\sigma_{w}^{2}+\frac{\lvert({\mathbf{e}}_{d}^{H}+\mathbf{v}^{H}{\mathbf{E}_{\mathrm{Z}}})(\hat{\mathbf{h}}_{d}+{\hat{\mathbf{Z}}}^{H}\mathbf{v})\rvert^2}{\lVert\hat{\mathbf{h}}_{d}^{H}+\mathbf{v}^{H}{\hat{\mathbf{Z}}}\rVert^2}}\right)\vline\hat{\mathbf{h}}_{d},\hat{\mathbf{Z}}\right\rbrace.
\end{align}

 The ergodic rate in (\ref{6}) contains nested  integrals over both the channel fading and the CSI estimation errors, which is mathematically intractable. To proceed, we resort to first presenting a tight approximation of (\ref{6}).
\begin{Theorem}
	\label{theorem}
Given an arbitrary  reflecting beamforming design $\mathbf{v}$,	the ergodic achievable rate of the network is tightly approximated as
		\begin{equation}
	\label{0304}
\bar{C}=\log_{2}\left(1+
	P\frac{\left|\mathbf{v}^{H}\mathbf{J}\mathbf{v}\right|^{2}}{\mathbf{v}^{H}\mathbf{Q}\mathbf{v}}\right),
	\end{equation}
	where $\mathbf{J}\triangleq\frac{\gamma_{1}}{NL}+\xi MNL+\mathbf{C}$, $\mathbf{Q}\triangleq\frac{\sigma_{w}^{2}\gamma_{1}+\gamma_{2}}{NL}+(\sigma_{w}^{2}M+\gamma_{3})NL\xi+(\sigma_{w}^{2}+\gamma_{4})\mathbf{C}$, $\mathbf{C}$ is a constant matrix given in (\ref{C24}) with respect to the channel statistical and the LoS components, and $\gamma_{1}\triangleq(1+\xi)M$, $\gamma_{2}\triangleq\xi M+\xi^{2}M(M+1)$, $\gamma_{3}\triangleq\gamma_{1}+\xi{M}+\xi(M+1)MNL$, and $\gamma_{4}\triangleq\xi(1+NL)$ are constant system parameters.
\end{Theorem}
\begin{IEEEproof}
	See Appendix \ref{zhengmingtheorem}.
\end{IEEEproof}

For the distributed IRSs and considering limited bandwidth of controlling signals for the IRSs, we design the reflecting beamformers by  utilizing statistical CSI. By applying \textit{Theorem 1} and temporarily ignoring   the constant-magnitude constrants of $\mathbf{v}$,  the   problem of optimizing $\mathbf{v}$ can be formulated as:
\begin{alignat}{2}
\label{imperfect}
&\qquad&\underset{\mathbf{v}}{\text{maximize}} \quad 
\frac{\left|\mathbf{v}^{H}\mathbf{J}\mathbf{v}\right|^{2}}
{\mathbf{v}^{H}\mathbf{Q}\mathbf{v}} .
\end{alignat}

\begin{Lemma}
	The optimal solution of  problem (\ref{imperfect}) is given by solving the equality $(\mathbf{I}-\mathbf{Q}^{-1}\mathbf{J})\mathbf{v}=\mathbf{0}$.  
\end{Lemma}
\begin{IEEEproof}
	According to the positive definiteness of $\mathbf{Q}$, let $\mathbf{w}=\mathbf{Q}^{1/2}\mathbf{v}$ and $\mathbf{B}=\mathbf{Q}^{-1/2}\mathbf{J}\mathbf{Q}^{-1/2}$. 
	Then the objective of (\ref{imperfect}) becomes $\frac{(\mathbf{w}^{H}\mathbf{B}\mathbf{w})^{2}}{\mathbf{w}^{H}\mathbf{w}}$. 
	Observing that for the scalar $\mathbf{w}^{H}\mathbf{B}\mathbf{B}\mathbf{w}\in\mathbb{C}$, we have $(\mathbf{B}\mathbf{w}\mathbf{w}^{H}\mathbf{B})\mathbf{B}\mathbf{w}=(\mathbf{w}^{H}\mathbf{B}\mathbf{B}\mathbf{w})\mathbf{B}\mathbf{w},$ which implies that 
   $\mathbf{B}\mathbf{w}$ is an eigenvector of the matrix $\mathbf{B}\mathbf{w}\mathbf{w}^{H}\mathbf{B}$.
   Since  the matrix $\mathbf{B}\mathbf{w}\mathbf{w}^{H}\mathbf{B}$ is of rank one, it is obvious that  $\frac{(\mathbf{w}^{H}\mathbf{B}\mathbf{w})^{2}}
	{\mathbf{w}^{H}\mathbf{w}}=\frac{\mathbf{w}^{H}(\mathbf{B}\mathbf{w}\mathbf{w}^{H}\mathbf{B})\mathbf{w}}
	{\mathbf{w}^{H}\mathbf{w}}\leq\lambda_{\max}(\mathbf{B}\mathbf{w}\mathbf{w}^{H}\mathbf{B})$, where $\lambda_{\max}$ is the maximum  eigenvalue and the equality holds when $\mathbf{w}=\mathbf{B}\mathbf{w}$, {i.e.,} $(\mathbf{I}-\mathbf{Q}^{-1}\mathbf{J})\mathbf{v}=\mathbf{0}$. The proof is complete.
\end{IEEEproof}

 Once we obtain an optimal solution of (\ref{imperfect}), say $\mathbf{v}'$, from \textit{Lemma 1}, the desired design of $\mathbf{v}$ with unit-magnitude elements for the IRSs can be subsequently achieved by using similar vector projection techniques as in \cite{passzhang}-\cite{estzhang3}. As an alternative, we can directly extract the phase of $\mathbf{v}'$ to obtain $\mathbf{v}^{*}=\angle\mathbf{v}'$, which is computationally efficient  and also  gives  close-to-optimal performance in most cases \cite{estzhang3}.
\subsection{Ergodic Rate Analysis}
\label{3B}
Given the derived results in \textit{Theorem 1}, we show that an insightful closed-form solution can be obtained for the special case with perfect CSI under uncorrelated Rician channels, which is summarized in the following Lemma.
\begin{Lemma}
	\label{lemmaper}
	For an uncorrelated channel with  $M = 1$ and perfect CSI at the transmitter, the ergodic achievable rate in (\ref{0304}) with the proposed statistical reflection design is given by
	\begin{equation}
	\label{performence2}
	\begin{aligned}
	\bar{C}&=\log_{2}\left(1+\frac{P}{\sigma_{w}^{2}}(1+\Upsilon_{1})\right)\triangleq C_{1},
	\end{aligned}
	\end{equation} 
	when $\Upsilon_{1}=L^{2}\left(\sum_{n=1}^{N}\sqrt{\frac{\alpha_{n}\beta_{n}K_{1n}K_{2,n}}{(K_{1,n}+1)(K_{2,n}+1)}}\right)^{2}+L\sum_{n=1}^{N}\frac{\alpha_{n}\beta_{n}({K_{2,n}}+K_{1,n}+1)}{(K_{1,n}+1)(K_{2,n}+1)}$.
\end{Lemma}
\begin{IEEEproof}
See Appendix \ref{proflemma2}.
\end{IEEEproof}

\begin{Remark}
	\label{remark1}
The result in (\ref{performence2}) implies that  
 the rate in (\ref{performence2}) is solely determined  by the large-scale fading coefficients and Rician factors. If the distances between the user and IRSs and between the BS and IRSs increase,  equivalently with decreasing $\beta_{n}$, $\alpha_{n}$, $K_{1,n}$, and $K_{2,n}$,  the performance  deteriorates.  
 To shed light on the design of the distributed IRS system, we further consider the system performance under some special cases as follows.
\end{Remark} 

\textbf{Case 1 (Pure LoS Propagations): }  If both channels of BS-to-IRSs and IRSs-to-user are pure LoSs, i.e., $K_{1n}=\infty$ and $K_{2n}=\infty$, $\forall n=1,...,N$, the ergodic achievable rate in (\ref{performence2}) reduces to 
\begin{align}
\label{case1}
C^{\mathrm{LoS}}_{1}=\log_{2}\left(1+\frac{P}{\sigma_{w}^{2}}\left(1+L^{2}\left(\sum_{n=1}^{N}\sqrt{\alpha_{n}\beta_{n}}\right)^{2}+L\sum_{n=1}^{N}\alpha_{n}\beta_{n}\right)\right).
\end{align} 

The rate in (\ref{case1}) is proportional to $L^{2}$ as expected, but also depends on the large-scale fadings.\textcolor{black}{ When we set the normalized $\beta_{n}=\alpha_{n}=1$, it gives}
$
C^{\mathrm{LoS}}_{1}=\log_{2}\left(1+\frac{P}{\sigma_{w}^{2}}(1+L^{2}N^{2}+LN)\right).
$

\textbf{Case 2 (Pure Rayleigh Channels):} If both channels of BS-to-IRSs and IRSs-to-user are Rayleigh distributed, i.e., $K_{1,n}=0$ and $K_{2,n}=0$, $\forall n=1,...,N$, in a rich-scattering scenario,  the ergodic achievable rate reduces to 
\begin{align}
\label{case2}
C^{\mathrm{R}}_{1}&=\log_{2}\left(1+\frac{P}{\sigma_{w}^{2}}\left(1+L\sum_{n=1}^{N}\alpha_{n}\beta_{n}\right)\right)\nonumber\\&\leq\log_{2}\left(1+\frac{P}{\sigma_{w}^{2}}(1+LN)\right),
\end{align} where the inequality in (\ref{case2}) holds \textcolor{black}{when we  normalized  $\alpha_{n}=\beta_{n}=1$.}

\begin{Remark}
	\label{remark2}
	Comparing the above two cases, we observe that the achievable rate becomes proportional to  $L$ and $N$ when both $\mathbf{{H}}$ and $\mathbf{g}$ are Rayleigh fading. When the LoS paths exist as in Case 1, the deployment of multiple IRSs improves the achievable rate significantly. Compared to a centralized deployment, distributed IRSs obviously increase the possibility of the presence of LoS channels.
\end{Remark}

\textbf{Case 3 (Hybrid Propagation):} For a general case, we define that the pure-LoS set $\mathcal{R}$. If $i\in\mathcal{R}$, we have  BS-to-IRS $j$ path and IRS $j$-to-user path are both LoS ($K_{1,j}=\infty$ and  $K_{2,j}=\infty$, $j\in\mathcal{R}$), while the IRS  in $\bar{\mathcal{R}}=\{1,...,N\}\backslash\mathcal{R}$ only has one or zero LoS path ($K_{1,j}=\infty$ and  $K_{2,j}=0$ or $K_{1,j}=0$ and  $K_{2,j}=\infty$). Then, the ergodic achievable rate in  (\ref{performence2}) can be rewritten as
\begin{align}
\label{case4}
&C^{\mathrm{Hybrid}}\notag\\=&\log_{2}\left(1+\frac{P}{\sigma_{w}^{2}}\left(1+L^{2}\left(\sum_{j\in\mathcal{M}}\sqrt{\alpha_{j}\beta_{j}}\right)^{2}+L\sum_{n=1}^{N}\alpha_{n}\beta_{n}\right)\right)\nonumber\\\leq&\log_{2}\left(1+\frac{P}{\sigma_{w}^{2}}(1+L^{2}m^{2}+LN)\right),
\end{align}
where $m=|\mathcal{R}|$ is the cardinality of $\mathcal{R}$.
\begin{Remark}
	\label{remark3}
	The result in (\ref{case4}) implies that the performance benefits brought by the distributed IRSs, rely on the presence of LoS paths. In practice, if only a subset of IRSs can be selected to serve a user, the IRSs that have clear LoS paths in the vicinity of the user should be selected to ensure better performance.
\end{Remark}
\section{Simulation Results}
\label{section4}
\begin{figure*}[!t]
	\centering
	\subfloat[]{\includegraphics[width=2.4in]{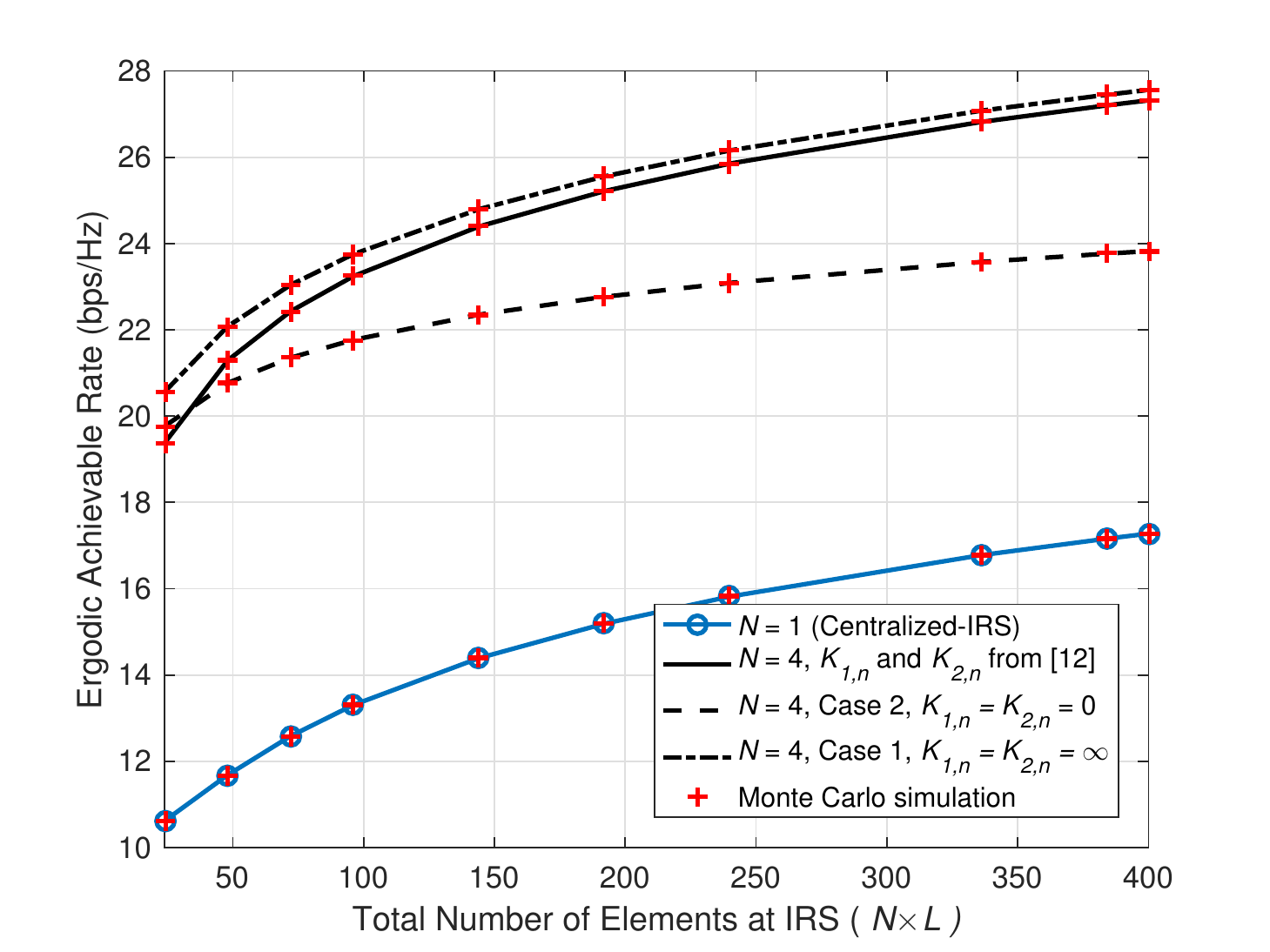}
		\label{fig:performance}}
	\subfloat[]{\includegraphics[width=2.4in]{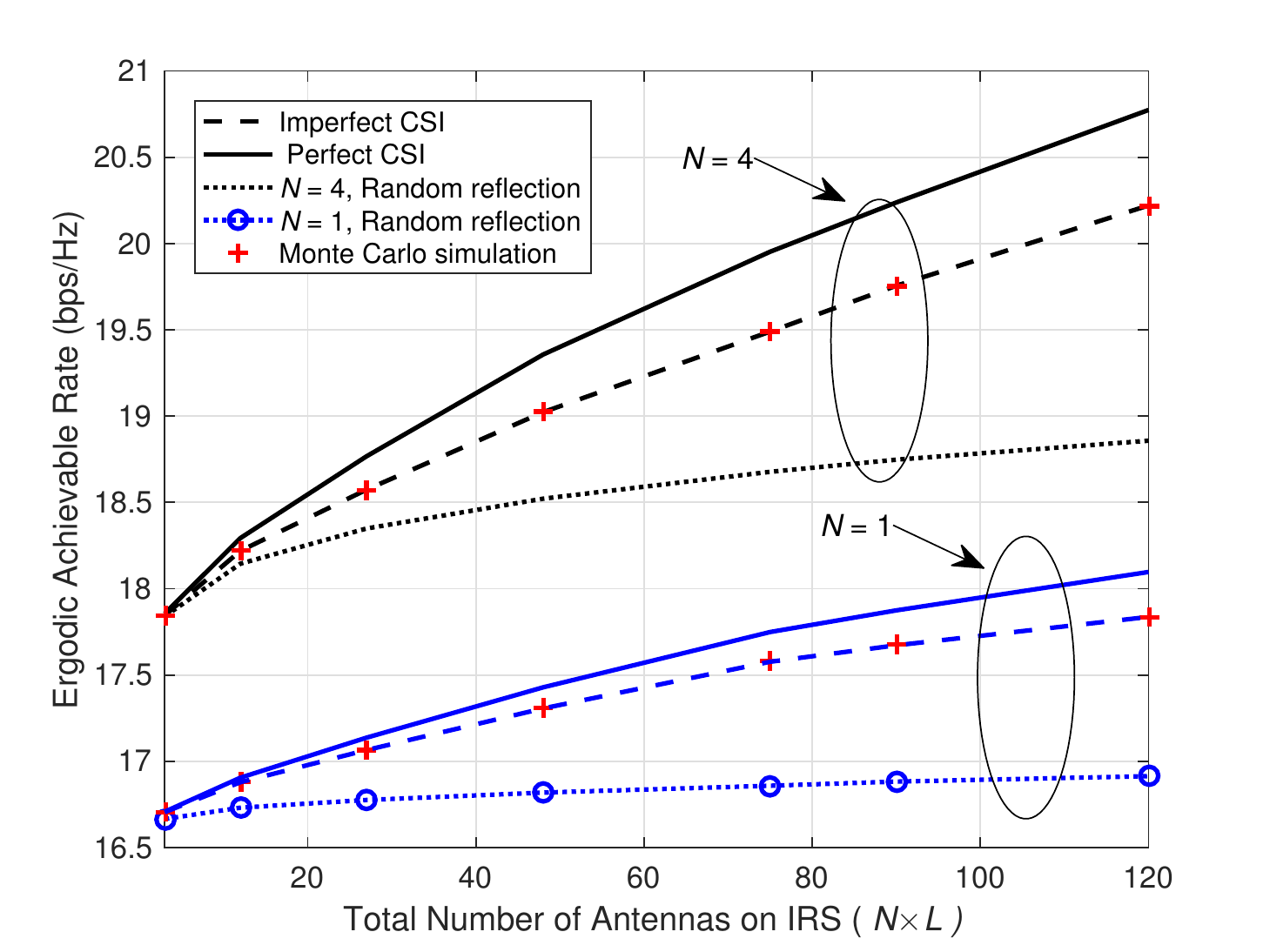}
		\label{fig:liscomm9}}
		\subfloat[]{\includegraphics[width=2.4in]{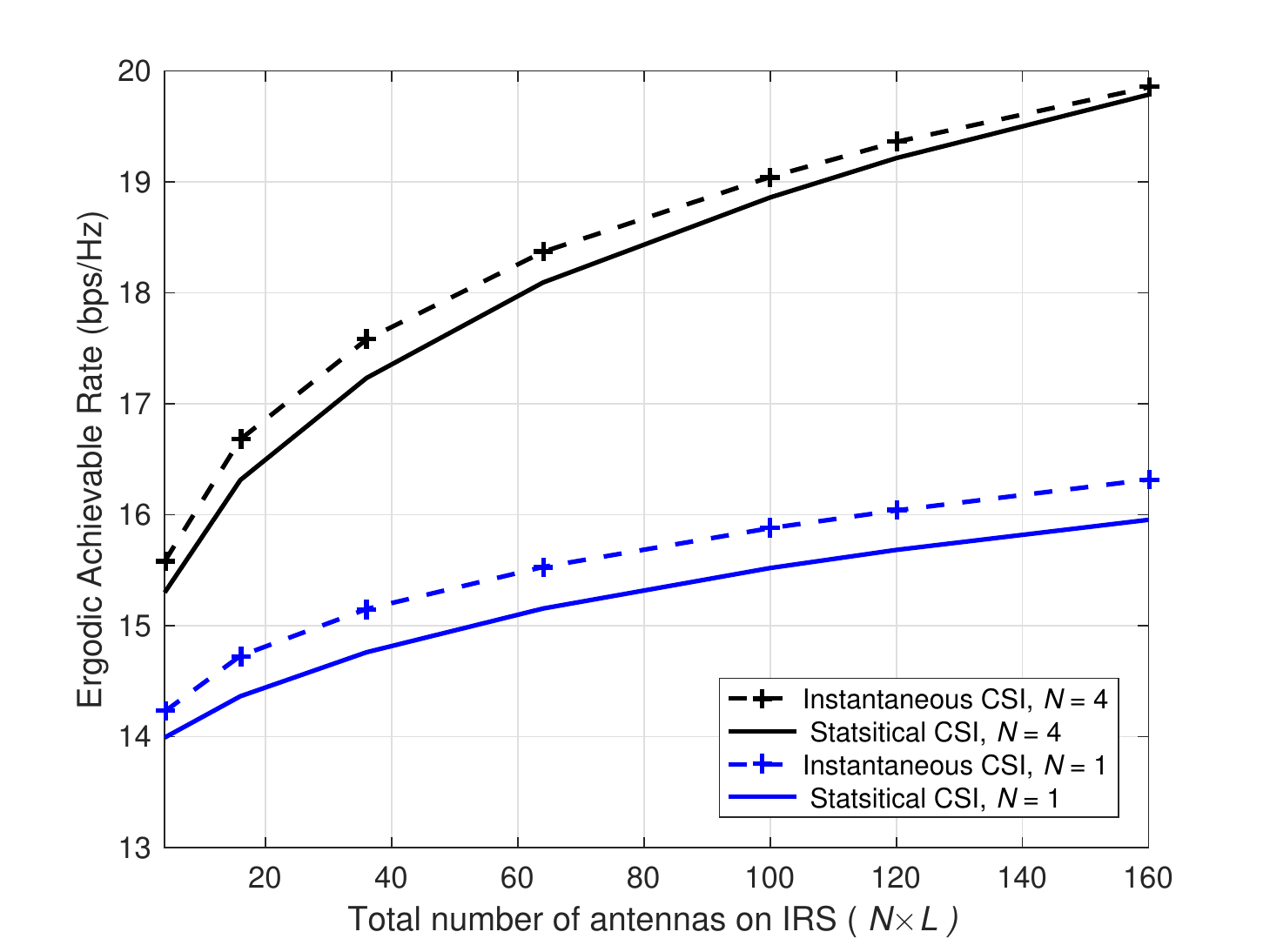}
		\label{fig:inscsi}}
	\caption{(a) Perfect CSI with $M = 1$. (b) Different CSI setups with $M = 9$. \textcolor{black}{(c) Perfect instantaneous CSI vs.  statistival CSI.}}
	\label{fig2}
\end{figure*}

\textcolor{black}{In this section, we provide numerical results to validate the
theoretical derivations. We set $T=10$, $\rho=20$ dB,  $\sigma_{w}^{2}=1$,
$P=20$ dB, $\beta_{n}=C_{0}{\left(\frac{d_{1,n}}{D_{0}}\right)^{-\alpha}}$, and $\alpha_{n}=C_{0}{\left(\frac{d_{2,n}}{D_{0}}\right)^{-\alpha}}$, where $C_{0}=10^{-3}$ is the path loss at the reference distance $D_{0}$ = 1 (m),  $\alpha=2.5$ is the pathloss exponent, $d_{1,n}\leq 10$ $\mathrm{m}$ and $d_{2,n}\leq 20$ $\mathrm{m}$ are, respectively, the distance between the BS and IRS $n$,  and the  distance between the user and IRS $n$. 
The statistical correlation matrix is
defined as in \cite{R}. The Rician coefficents are set as $K_{1,n}=10^{1.3-0.003d_{1,n}}$ and $K_{2,n}=10^{1.3-0.003d_{2,n}}$ \cite{jinshi}. The IRSs are uniformly distributed in the cell coverage and simulation
	results are averaged with 1000 fading channel realizations to ensure fairness. 
}

 Fig. \ref{fig:performance} cshows the ergodic achievable rate versus the total number of reflecting
 elements with perfect CSI. Solid lines correspond to analytical results under Rician fading channels while dotted markers correspond to 
 Case 1
  and Case 2. 
  The system with 4 distributed IRSs, $N=4$, outperforms the system with a centralized-IRS, i.e., $N=1$, with the same total number of reflecting elements $NL$. As $NL$ increases, the performance gap between the results in solid line and these in Case 1 diminishes, which agrees with the results in \cite{yuhan},  where the gap is caused by  the coefficients $\frac{K_{1n}K_{2,n}}{(K_{1,n}+1)(K_{2,n}+1)}<1$ and $\frac{K_{1n}+K_{2,n}+1}{(K_{1,n}+1)(K_{2,n}+1)}<1$  in (\ref{performence2}) become less dominated. Besides, the gap between Case 1 and Case 2 enlarges since the SNR loss $N^{2}L^{2}\frac{P}{\sigma_{w}^{2}}$ in Case 2 is proportional to $L^{2}$, as predicted in \textit{Remark \ref{remark2}} and \textit{Remark \ref{remark3}}.

Fig. \ref{fig:liscomm9} shows the ergodic achievable rate versus the total number of reflecting
elements. The analytical
result in (\ref{0304}) tightly matches with the numerical results via Monte Carlo simulation. Besides, the considered distributed IRSs outperform the  centralized-IRS.
 Compared with random reflection  $\Phi_{n}$, the results of our proposed reflection design are considerably higher due to the optimized phase shifts at IRSs. \textcolor{black}{The approximation in our analysis is  tight mainly because the dimensions of the matrices and vectors, i.e., $N\times L$, are  very large,  which satisfies the assumption of large antenna array for using the mathematical approximation.}
 
\textcolor{black}{ For fair comparison, we compare the performance 	gains between instantaneous CSI and statistical CSI under the assumption of no channel estimation error for fairnes. With perfect instantaneous CSI, we exploit the solution in \cite{perloss} to design the reflection vector in a SISO system.  As shown in this new figure results, i.e., Fig. \ref{fig:inscsi}, we can find that the performance loss between the reflection design with instantaneous CSI and that with perfect statistical CSI are marginal.}

\section{Conclusion}
\label{section5}
 We proposed a low-complexity reflection design for IRSs  with statistical CSI. The derived ergodic achievable
rate can accurately characterize the system performance of the distributed IRSs system, which outperforms the one with a 
centralized IRS
with the same number of IRS elements.


%
\appendices
\section{Proof of Theorem \ref{theorem}}
\label{zhengmingtheorem}
In order to facilitate the analysis, we apply the result  in \cite[Lemma 1]{jinshi} which proves that the approximation of $\mathbb{E}\{\log_{2}(1+\frac{X}{Y})\}\approx\log_{2}\left(1+\frac{\mathbb{E}\{X\}}{\mathbb{E}\{Y\}}\right)$ is accurate enough  for  massive MIMO, e.g., with a large number of antennas $NL$ and $M$.
Hence, we can approximate the rate in (\ref{6}) as
\begin{equation}
\label{13}
\bar{C}\approx\text{log}_{2}\left(1+P\frac{\mathbb{E}\{\lVert\hat{\mathbf{h}}_{d}^{H}+\mathbf{v}^{H}{\hat{\mathbf{Z}}}\rVert^2\}}{\sigma_{w}^{2}+\frac{\mathbb{E}\{\lvert({\mathbf{e}}_{d}^{H}+\mathbf{v}^{H}{\mathbf{E}_{\mathrm{Z}}})(\hat{\mathbf{h}}_{d}+{\hat{\mathbf{Z}}}^{H}\mathbf{v})\rvert^2\}}{\mathbb{E}\{\lVert\hat{\mathbf{h}}_{d}^{H}+\mathbf{v}^{H}{\hat{\mathbf{Z}}}\rVert^2\}}}\right).
\end{equation}
To proceed, we first derive the  numerator in  (\ref{13}) as
\begin{align}
\label{fenzi}
&\mathbb{E}\{\lVert\hat{\mathbf{h}}_{d}^{H}+\mathbf{v}^{H}{\hat{\mathbf{Z}}}\rVert^2\}\notag\\
&\overset{(a)}{=}\mathbb{E}\{\hat{\mathbf{h}}_{d}^{H}\hat{\mathbf{h}}_{d}\}+\mathbb{E}\{\mathbf{v}^{H}{\hat{\mathbf{Z}}}\hat{\mathbf{Z}}^{H}\mathbf{v}\}\notag\\
&=\mathbb{E}\{\mathbf{h}_{d}^{H}\mathbf{h}_{d}\}+\mathbb{E}\{\mathbf{e}_{d}^{H}{\mathbf{e}}_{d}\}+\mathbf{v}^{H}\mathbb{E}\{\mathbf{E}_{\mathrm{Z}}{\mathbf{E}_{\mathrm{Z}}}^{H}\}\mathbf{v}+\mathbf{v}^{H}\mathbb{E}\{\mathbf{Z}\mathbf{Z}^{H}\}\mathbf{v}
\notag\\&\overset{(b)}{=}M+M\xi+MNL\xi+\mathbf{v}^{H}\mathbb{E}\{\mathbf{Z}\mathbf{Z}^{H}\}\mathbf{v},
\end{align} 
where $(a)$ exploits the independence between $\hat{\mathbf{h}}_{d}$ and $\hat{\mathbf{Z}}$ since $\mathbf{Z}$, $\mathbf{E}_{\mathrm{Z}}$,   $\mathbf{h}_{d}$, and ${\mathbf{e}}_{d}$ are independent, while $(b)$ is due to $\mathbf{h}_{d}\sim\mathcal{CN}(\mathbf{0}, \mathbf{I}_{M})$, ${\mathbf{e}}_{d}\sim\mathcal{CN}(\mathbf{0}, \xi\mathbf{I}_{M})$, and ${\mathbf{E}_{\mathrm{Z}}}\sim\mathcal{CN}(\mathbf{0}, \xi\mathbf{I}_{NL}\otimes\mathbf{I}_{M})$ for large $M$, $NL$.
Then, we evaluate $\mathbf{v}^{H}\mathbb{E}\{\mathbf{Z}\mathbf{Z}^{H}\}\mathbf{v}$ in (\ref{fenzi}) as follows
\begin{align}
\label{GG}
&\mathbf{v}^{H}\mathbb{E}\{\mathbf{Z}\mathbf{Z}^{H}\}\mathbf{v}\notag\\
&\overset{(a)}{=}\mathbf{v}^{H}(\bar{\mathbf{G}}\bar{\mathbf{{H}}}\bar{\mathbf{H}}^{H}\bar{\mathbf{G}}^{H}+\bar{\mathbf{G}}\mathbf{K}_{1}\mathbb{E}\{\widetilde{\mathbf{H}}\widetilde{\mathbf{H}}^{H}\}\mathbf{K}_{1}\bar{\mathbf{G}}^{H}\nonumber\\&+\mathbf{K}_{2}\mathbb{E}\{\widetilde{\mathbf{G}}\bar{\mathbf{H}}\bar{\mathbf{H}}^{H}\widetilde{\mathbf{G}}^{H}\}\mathbf{K}_{2}+\mathbf{K}_{2}\mathbb{E}\{\widetilde{\mathbf{G}}\mathbf{K}_{1}\widetilde{\mathbf{H}}\widetilde{\mathbf{H}}^{H}\mathbf{K}_{1}\widetilde{\mathbf{G}}^{H}\}\mathbf{K}_{2})\mathbf{v}
\nonumber\\&\overset{(b)}{=}\mathbf{v}^{H}(\bar{\mathbf{G}}\bar{\mathbf{{H}}}\bar{\mathbf{H}}^{H}\bar{\mathbf{G}}^{H}+M\bar{\mathbf{G}}\mathbf{K}_{1}\mathbf{R}\mathbf{K}_{1}\bar{\mathbf{G}}^{H})\mathbf{v}\nonumber\\&+\text{tr}\{\mathbf{K}_{2}\bar{\mathbf{H}}\bar{\mathbf{H}}^{H}\mathbf{K}_{2}\}+M\text{tr}\{\mathbf{K}_{2}\mathbf{K}_{1}\mathbf{R}\mathbf{K}_{1}\mathbf{K}_{2}\},
\end{align}
where $(a)$ exploits the independence between $\widetilde{\mathbf{G}}$ and $\widetilde{\mathbf{H}}$ and the fact that $\bar{\mathbf{G}}$ and $\bar{\mathbf{{H}}}$ are constants, and $(b)$ follows by $\widetilde{\mathbf{H}}\sim\mathcal{CN}(\mathbf{0}, \mathbf{R}\otimes\mathbf{I}_{M})$ and the diagonality of $\widetilde{\mathbf{G}}$  whose entries are independent and identically distributed (i.i.d.) Gaussian distributed with zero mean and unit variance.

Substituting (\ref{GG}) into equation (\ref{fenzi}) yields
\begin{align}
\label{fenziresult}
&\mathbb{E}\{\lVert\hat{\mathbf{h}}_{d}^{H}+\mathbf{v}^{H}{\hat{\mathbf{Z}}}\rVert^2\}\\&=(1+\xi)M+MNL\xi+\mathbf{v}^{H}(M\mathbf{K}_{2}\text{tr}\{\mathbf{K}_{1}\mathbf{R}\mathbf{K}_{1}\}\mathbf{K}_{2}\notag\\&+M\bar{\mathbf{G}}\mathbf{K}_{1}\mathbf{R}\mathbf{K}_{1}\bar{\mathbf{G}}^{H}+\mathbf{K}_{2}\text{tr}\{\bar{\mathbf{H}}\bar{\mathbf{H}}^{H}\}\mathbf{K}_{2}+\bar{\mathbf{G}}\bar{\mathbf{{H}}}\bar{\mathbf{H}}^{H}\bar{\mathbf{G}}^{H})\mathbf{v}.\notag
\end{align}
For the denominator in (\ref{13}), we have
\begin{equation}
\label{fenmu}
\begin{aligned}
&\mathbb{E}\{\lvert({\mathbf{e}}_{d}^{H}+\mathbf{v}^{H}{\mathbf{E}_{\mathrm{Z}}})(\hat{\mathbf{h}}_{d}+{\hat{\mathbf{Z}}}^{H}\mathbf{v})\rvert^2\}\\
&=\mathbb{E}\{{\mathbf{e}}_{d}^{H}\hat{\mathbf{h}}_{d}\hat{\mathbf{h}}_{d}^{H}{\mathbf{e}}_{d}\}+\mathbb{E}\{\mathbf{v}^{H}{\mathbf{E}_{\mathrm{Z}}}\hat{\mathbf{h}}_{d}\hat{\mathbf{h}}_{d}^{H}{\mathbf{E}_{\mathrm{Z}}}^{H}\mathbf{v}\}
\\&+\mathbb{E}\{{\mathbf{e}}_{d}^{H}\hat{\mathbf{Z}}^{H}\mathbf{v}\mathbf{v}^{H}\hat{\mathbf{Z}}{\mathbf{e}}_{d}\}+\mathbb{E}\{\mathbf{v}^{H}{\mathbf{E}_{\mathrm{Z}}}\hat{\mathbf{Z}}^{H}\mathbf{v}\mathbf{v}^{H}\hat{\mathbf{Z}}{\mathbf{E}_{\mathrm{Z}}}^{H}\mathbf{v}\}.
\end{aligned}
\end{equation}

The four expectation terms in (\ref{fenmu}) can be, respectively, calculated as
\begin{align}
\label{fenmu3}
&\mathbb{E}\{{\mathbf{e}}_{d}^{H}\hat{\mathbf{Z}}^{H}\mathbf{v}\mathbf{v}^{H}\hat{\mathbf{Z}}{\mathbf{e}}_{d}\}
=\mathbf{v}^{H}\mathbb{E}\{\hat{\mathbf{Z}}{\mathbf{e}}_{d}{\mathbf{e}}_{d}^{H}\hat{\mathbf{Z}}^{H}\}\mathbf{v}
\notag\\&=\xi{\mathbf{v}^{H}\mathbb{E}\{\hat{\mathbf{Z}}\hat{\mathbf{Z}}^{H}\}\mathbf{v}},\\
\label{wis}
&\mathbb{E}\{\mathbf{v}^{H}\mathbf{E}_{Z}\hat{\mathbf{Z}}^{H}\mathbf{v}\mathbf{v}^{H}\hat{\mathbf{Z}}\mathbf{E}_{Z}^{H}\mathbf{v}\}&\notag\\
&=\mathbf{v}^{H}\mathbb{E}\{{\mathbf{Z}}\mathbb{E}\{\mathbf{E}_{\mathrm{Z}}^{H}\mathbf{v}\mathbf{v}^{H}\mathbf{E}_{\mathrm{Z}}{\mathbf{Z}}^{H}\}\}\mathbf{v}+\mathbb{E}\{|\mathbf{v}^{H}\mathbf{E}_{\mathrm{Z}}\mathbf{E}_{\mathrm{Z}}^{H}\mathbf{v}|^{2}\}&\notag\\
&=\xi NL\mathbf{v}^{H}\mathbb{E}\{\mathbf{Z}\mathbf{Z}\}\mathbf{v}+\xi^{2}M(M+1)N^{2}L^{2},\\
\label{fenmu1}
&\mathbb{E}\{{\mathbf{e}}_{d}^{H}\hat{\mathbf{h}}_{d}\hat{\mathbf{h}}_{d}^{H}{\mathbf{e}}_{d}\}=\mathbb{E}\{{\mathbf{e}}_{d}^{H}{\mathbf{h}}_{d}{\mathbf{h}}_{d}^{H}{\mathbf{e}}_{d}\}+\mathbb{E}\{{\mathbf{e}}_{d}^{H}{\mathbf{e}}_{d}{\mathbf{e}}_{d}^{H}{\mathbf{e}}_{d}\}\notag\\&=\xi M+\xi^{2}M(M+1),
\\
&\text{and}\notag\\
\label{fenmu2}
&\mathbf{v}^{H}\mathbb{E}\{{\mathbf{E}_{\mathrm{Z}}}\hat{\mathbf{h}}_{d}\hat{\mathbf{h}}_{d}^{H}{\mathbf{E}_{\mathrm{Z}}}^{H}\}\mathbf{v}\notag\\&=\mathbf{v}^{H}\mathbb{E}\{{\mathbf{E}_{\mathrm{Z}}}{\mathbf{h}}_{d}{\mathbf{h}}_{d}^{H}{\mathbf{E}_{\mathrm{Z}}}^{H}\}\mathbf{v}+\mathbf{v}^{H}\mathbb{E}\{{\mathbf{E}_{\mathrm{Z}}}{\mathbf{e}}_{d}{\mathbf{e}}_{d}^{H}{\mathbf{E}_{\mathrm{Z}}}^{H}\}\mathbf{v}\notag\\&=MNL\xi+{M}NL\xi^{2}=(1+\xi)\xi MNL,
\end{align}
where the last equality in (\ref{wis}) holds because $\mathbf{v}^{H}\mathbf{Z}\sim\mathcal{CN}(\mathbf{0},\xi NL\mathbf{I}_{M})$ and we obtain $\mathbb{E}\{{w}(\xi NL)^{-1}{w}\}=M(M+1)NL\xi$, i.e., $\mathbb{E}\{|\mathbf{v}^{H}\mathbf{E}_{\mathrm{Z}}\mathbf{E}_{\mathrm{Z}}^{H}\mathbf{v}|^{2}\}=\xi^{2}M(M+1)N^{2}L^{2}$ as ${w}\triangleq\mathbf{v}^{H}\mathbf{E}_{\mathrm{Z}}\mathbf{E}_{\mathrm{Z}}^{H}\mathbf{v}$ is an uncorrelated central Wishart random  variable, commonly denoted as ${w}\sim\mathcal{CW}_{1}(M,NL\xi)$ \cite{wishart}.

Substituting (\ref{fenmu1}) and (\ref{wis}) into (\ref{GG}) and using (\ref{fenzi}), we have
\begin{align}
\label{C24}
&\mathbb{E}\{\lvert({\mathbf{e}}_{d}^{H}+\mathbf{v}^{H}\mathbf{E}_{\mathrm{Z}})(\hat{\mathbf{h}}_{d}+{\hat{\mathbf{Z}}}^{H}\mathbf{v})\rvert^2\}\notag\\&=\xi M+\xi^{2}M(M+1)+(1+\xi)MNL\xi+\xi^{2}MNL\notag\\&+(\xi+NL\xi)\mathbf{v}^{H}\mathbf{C}\mathbf{v}+M(M+1)N^{2}L^{2}\xi^{2},
\end{align}
where $\mathbf{C}=\bar{\mathbf{G}}\bar{\mathbf{{H}}}\bar{\mathbf{H}}^{H}\bar{\mathbf{G}}^{H}+M\bar{\mathbf{G}}\mathbf{K}_{1}\mathbf{R}\mathbf{K}_{1}\bar{\mathbf{G}}^{H}+\frac{1}{NL}\text{tr}\{\mathbf{K}_{2}\bar{\mathbf{H}}\bar{\mathbf{H}}^{H}\mathbf{K}_{2}\}+\frac{M}{NL}\text{tr}\{\mathbf{K}_{2}\mathbf{K}_{1}\mathbf{R}\mathbf{K}_{1}\mathbf{K}_{2}\}$.
 Now by substituting (\ref{fenziresult}) and (\ref{C24}) into (\ref{13}), we complete the proof.
\section{Proof of Lemma 2}
\label{proflemma2}
The ergodic rate in (\ref{0304}) with perfect CSI for beamforming design, i.e., $\xi=0$, can be rewritten as  
$
\bar{C}=\log_{2}\left(1+\frac{P}{\sigma_{w}^{2}}(M+\mathbf{v}^{H}\mathbf{C}\mathbf{v})\right).
$

Considering the term $s\triangleq\mathbf{v}^{H}\mathbf{C}\mathbf{v}$, we have
\begin{align}
\label{performancea}
s=&\left\Vert\mathbf{v}^{H}\bar{\mathbf{G}}\bar{\mathbf{{H}}}\right\Vert^{2}+\text{tr}\{\bar{\mathbf{G}}\mathbf{K}_{1}\mathbf{K}_{1}\bar{\mathbf{G}}^{H}\}\notag\\+&	\text{tr}\{\bar{\mathbf{H}}\bar{\mathbf{H}}^{H}\mathbf{K}_{2}\mathbf{K}_{2}\}+\text{tr}\{\mathbf{K}_{2}\mathbf{K}_{1}\mathbf{K}_{1}\mathbf{K}_{2}\}\notag\\
=&\left(\sum_{n=1}^{N}\left(\sqrt{\frac{\alpha_{n}\beta_{n}K_{1,n}K_{2,n}}{(K_{1,n}+1)(K_{2,n}+1)}}\sum_{i=1}^{L}e^{j\phi_{ni}+j\theta_{ni}^{g}+j\theta_{ni}^{H}}\right)\right)^{2}\notag\\+&\sum_{n=1}^{N}\left(\frac{L\alpha_{n}\beta_{n}K_{2,n}}{(K_{1,n}+1)(K_{2,n}+1)}+\frac{L\alpha_{n}\beta_{n}K_{1,n}}{(K_{1,n}+1)(K_{2,n}+1)}\right)\notag\\+&\sum_{n=1}^{N}\left(\frac{L\alpha_{n}\beta_{n}}{(K_{1,n}+1)(K_{2,n}+1)} \right)\notag\\
\leq& L^{2}\left(\sum_{n=1}^{N}\sqrt{\frac{\alpha_{n}\beta_{n}K_{1,n}K_{2,n}}{(K_{1,n}+1)(K_{2,n}+1)}}\right)^{2}
\notag\\+&L\sum_{n=1}^{N}\frac{\alpha_{n}\beta_{n}(K_{1,n}+{K_{2,n}}+1)}{(K_{1,n}+1)(K_{2,n}+1)},
\end{align}
where $\theta_{ni}^{H}=\angle\bar{\mathbf{{H}}}_{ni}$ and $\theta_{ni}^{g}=\angle\bar{\mathbf{g}}_{ni}$ are the angles of the small-scale fading coefficients of $\bar{\mathbf{{H}}}$ and $\bar{\mathbf{g}}$, respectively. The inequality in (\ref{performancea}) holds with equality if and only if $\mathbf{v}$ is designed by:
$	\phi_{ni}=-(\theta_{ni}^{g}+\theta_{ni}^{H})$, $n=\{1,...,N\}$, and $i=\{1,...,L\}$.
 This completes the proof.



\ifCLASSOPTIONcaptionsoff
  \newpage
\fi


\begin{thebibliography}{1}
  
\bibitem{R}
 D. W. K. Ng, E. S. Lo, and R. Schober, ``Energy-efficient resource allocation in OFDMA systems with large numbers of base station antennas,"  \emph{IEEE Trans. Wireless Commun.}, vol. 11, no. 9, pp. 3292$-$3304, Sep. 2012.
  \bibitem{xujindan}S. Zhou, W. Xu, K. Wang, M. Di Renzo and M. Alouini, "Spectral and energy efficiency of IRS-assisted MISO communication with hardware ompairments,"  \emph{IEEE Wireless Commun. Lett.}, pp. 1$-$1, Apr. 2020
  \bibitem{passzhang} Q. Wu and R. Zhang, ``Intelligent reflecting surface enhanced wireless network: joint active and passive beamforming design," in \emph{Proc. IEEE GLOBECOM}, United Arab Emirates, pp. 1$-$6, Oct. 2018.
  \bibitem{yuhan} Y. Han, W. Tang, S. Jin, C. Wen, and X. Ma, ``Large intelligent surface-assisted wireless communication exploiting statistical CSI," \emph{IEEE Trans. Veh. Tech.}, vol. 68, no. 8, pp. 8238$-$8242, Aug. 2019.
    \bibitem{estzhang3} D. Mishra and H. Johansson, ``Channel estimation and low-complexity beamforming design for passive intelligent surface-assisted MISO wireless energy transfer," in \emph{Proc. IEEE ICASSP}, May 2019.
  \bibitem{estimationzhanf}S. Zhou, W. Xu, K. Wang, C. Pan, M. Alouini, and A. Nallanathan, ``Ergodic rate analysis of cooperative ambient backscatter communication,"  \emph{IEEE Wireless Commun. Lett.}, vol. 8, no. 6, pp. 1679$-$1682, Dec. 2019.


\bibitem{ref2}J. Lyu and R. Zhang, ``Hybrid Active/Passive Wireless Network Aided by Intelligent Reflecting Surface: System Modeling and Performance Analysis," [Online]. Available: https://arxiv.org/abs/2004.13318. 
\bibitem{spatial}J. Lyu and R. Zhang, ``Spatial throughput characterization for intelligent reflecting surface aided multiuser system,'' \emph{IEEE Wireless Commun. Lett.}, vol. 9, no. 6, pp. 834$-$838, Jun. 2020.

\bibitem{jinshi}
Q. Zhang, S. Jin, and K-K. Wong, ``Power scaling of uplink massive MIMO systems
with arbitrary-rank channel means," \emph{ IEEE J.  Sel. Topics  Signal Process.}, vol. 8, no. 5, pp. 966$-$981. Oct. 2014.
\bibitem{CSIMIMO}T. C. Mai, H. Q. Ngo, and T. Q. Duong, ``Downlink spectral efficiency of cell-free massive
MIMO systems with multi-antenna users," in \emph{Proc. IEEE GlobalSIP,} USA, 2018, pp. 828$-$832.
\bibitem{error}O. Raeesi and A. Gokceoglu, ``Performance analysis of multi-user massive MIMO
downlink under channel non-reciprocity
and imperfect CSI,"   \emph{IEEE Trans. Commun.},  vol. 66, no. 6, pp. 2456$-$2471, Jun. 2018.


\bibitem{perloss} E. G. Larsson, ``Intelligent reflecting surface versus decode-and-forward: How large surfaces are needed to beat relaying?", \emph{ IEEE Wireless Commun. Lett.,} vol. 9, no. 2, pp. 244$-$248, Feb. 2020.

\bibitem{wishart} K. N. Arjun, Daya,  and K. Gupta, ``Expectations of functions of complex wishart matrix," Acta Appl Math, Mar. 2011.



\end{thebibliography}
\end{document}